\documentclass{iopjournal}
\usepackage{amsmath}
\usepackage[dvipsnames]{xcolor}
\usepackage{physics}
\usepackage{mathtools}
\usepackage{amssymb}
\usepackage{cite}

\usepackage{caption}
\usepackage{subcaption}
\usepackage{graphicx}
\usepackage{anyfontsize}
\usepackage{mathrsfs}

\usepackage{tikz}
\usetikzlibrary{positioning}
\usepackage{tkz-berge}
\usetikzlibrary{arrows}
\usetikzlibrary{graphs,graphs.standard,fit,shapes.geometric,calc,hobby}
\newcommand{\cb}[1]{\textcolor{black}{#1}}

\begin{document}
\articletype{Research Paper}

\title{Routing Qubits on Noisy Networks}
\author{Claudia Benedetti$^{1,*}$\orcid{0000-0002-8112-4907}, Giovanni Ragazzi $^2$\orcid{0009-0007-0980-4741}, Simone Cavazzoni$^2$\orcid{0000-0001-7975-5059}, Paolo Bordone$^{2,3}$\orcid{0000-0002-4313-0732} and Matteo G. A. Paris $^{1,4}$\orcid{0000-0001-7523-7289}}

\affil{$^1$Quantum Technology Lab, Dipartimento di Fisica {\em Aldo Pontremoli}, Universit\`{a} di Milano, I-20133 Milano, Italy}
\affil{$^2$Dipartimento di Scienze Fisiche, Informatiche e Matematiche,  Universit\`{a} di Modena e Reggio Emilia, I-41125 Modena, Italy}
\affil{$^3$Centro S3, CNR-Istituto di Nanoscienze, I-41125 Modena, Italy}
\affil{$^4$INFN, Sezione di Milano, I-20133 Milano, Italy}

\affil{$^*$Author to whom any correspondence should be addressed.}
\email{claudia.benedetti@unimi.it}


\begin{abstract}
Robust quantum routing is essential for scalable quantum technologies. This paper investigates 
the resilience of routing protocols in network architectures designed for perfect, high-fidelity transfer of both classical and quantum information under ideal conditions. We encode information 
in the position of a quantum walker on a graph, modelling the routing of a generic qubit state 
from a single input to multiple (orthogonal) outputs. We analyse and assess routing performance 
in various regimes, evaluating their robustness against static and dynamical noise.
\end{abstract}

\section{Introduction}
\label{sec:I}
Quantum routing is a fundamental protocol for directing  quantum information encoded in a general quantum state across physical systems \cite{zueco2009quantum,Ahumada_2019,Bottarelli_2023}. It consists in applying controlled Hamiltonian operations that select specific communication paths without disturbing the quantum states being transmitted. This capability is becoming increasingly relevant, as nearly all quantum information protocols, from computation to networking, require efficient and robust methods to exchange information both within quantum devices and over long distances. The field is receiving attention in several directions. On a hardware level, routing can be implemented in noisy, near-term quantum devices integrated with error correction schemes \cite{shi2023quantum}. Alternatively, a router architecture can be constructed from quantum memories connected via a photonic switchboard to manage entanglement distribution across quantum networks \cite{lee2022quantum}. Furthermore, machine learning techniques are being employed to design architectures that optimize routing paths for maximum throughput \cite{le2022dqra}. Experimentally, a wide variety of platforms have demonstrated quantum routing. Successful proposals and implementations have been based on interferometric approaches  \cite{yuan2015experimental},  superconducting quantum circuits \cite{wang2021experimental}, optical systems \cite{yan2014single,bartkiewicz2018implementation, ham2004experimental,zhou2013quantum}, among others \cite{cao2017implementation,nannicini2022optimal,wagner2023improving,huang2024efficient,palaiodimopoulos2024chiral}. On the other hand, network-theoretical tools often provide a distinct advantage by significantly reducing the number of required measurements compared to usual methods applied in repeater schemes \cite{hahn2019quantum}. 

Quantum spin chains provide a well-established platform for implementing quantum state transfer \cite{karbach2005spin,gualdi2008perfect,yao2011robust,vinet2012construct} and routing \cite{PhysRevLett.106.020503,paganelli2013routing}. Indeed, in the single-output setting, this problem is typically referred to as quantum state transfer and has been widely investigated in those systems  \cite{bose2003quantum,zwick2012spin}. In spin models, if the Hamiltonian conserves the number of excitations, the problem can be reformulated as a continuous-time quantum walk (CTQW), by restricting the analysis to the single-excitation subspace \cite{kay2010perfect}. The goal, in this case, is to transfer an arbitrary superposition state from one spin to another.  Within this setting, it is known how to choose the edge weights to achieve perfect state transfer \cite{christandl2004perfect}. Similarly, the task of quantum routing can be analysed in  the single excitation subspace, i.e. in terms of CTQWs on the underlying network, thus offering a convenient representation for studying quantum transfer schemes in networked quantum systems. Single-particle CTQWs describe the coherent evolution of an excitation over a discrete set of positions whose connectivity defines the underlying graph \cite{mulken2011continuous,fahri98}. Extending the CTQW concept to Hamiltonians with complex-valued couplings introduces directionality into the dynamics, giving rise to chiral  quantum walks \cite{Lu2016, Zimboras13,frigerio2021generalized}. QW inherent quantum properties, like self-interference, together with chirality enables routing of quantum and classical information \cite{gao2023demonstration,sauglam2023entanglement}. Moreover, chiral QWs have been employed in quantum transport and state transfer schemes \cite{Zimboras13,todtly16,Khalique2021,acuaviva25,Yu23,finocchiaro2025optimal} as well as for    quantum algorithms \cite{Wong2015,frigerio22b}. Progress in the study  of chiral QW has lead to networks models that theoretically support pretty good \cite{ragazzi2025scalable} and  perfect routing through chiral control \cite{cavazzoni2025perfect}. In this paper, we investigate the robustness of quantum routing protocols implemented in an  architecture that, under ideal conditions, support perfect routing of both classical and quantum information (i.e., with unit routing fidelity). Our goal is to assess how imperfections and noise  affect the routing performances in possible implementations.

This paper is organized as follows. In Section \ref{sec:M} we review the theoretical framework 
underlying CTQWs and quantum routing, and introduce a network topology supporting perfect quantum routing. 
We then introduce two paradigmatic kinds of noise, modeled by static von Mises or Gaussian error 
distributions, and dynamic Ornstein-Uhlenbeck processes. In Section \ref{sec:R}, we discuss 
the robustness of routing against noise, identifying noise regimes in which the  routing fidelity remains sufficiently high to enable a reliable transmission of quantum information. Finally, in Sections \ref{sec:D} and \ref{sec:C}, we provide a summary of our findings and discuss their implications for the transmission of quantum information.
\section{Method}
\label{sec:M}
In this section, we first review the fundamental concepts of graph theory together with the basis of routing procedures, which allows us to define the network structure and its associated quantum dynamics. In this framework, a router is a quantum network with one input region associated with $\ket{\psi^{\rm{in}}}$ and $n$ possible outputs associated with $\{ \ket{\psi^{\rm{out}}}_{\xi=1,...,n} \}$. We then analyse the impact of multiple noise sources on the ideal routing protocol.
\subsection{Quantum walks on  networks}
The transfer of the information encoded onto a quantum state across a discrete structure can be modelled through a continuous-time quantum walk on a graph, in which the time evolution of the system takes place in a position Hilbert space $\mathscr{H} = \operatorname{span}\{ \vert x \rangle \}$. The states $\vert x \rangle$ are the sites of a $\mathcal{N}$-dimensional discrete network associated with a graph $\mathcal{G} \left( \mathcal{V},\mathcal{E} \right)$ with vertices (or nodes) $\mathcal{V}$ and edges $\mathcal{E}$. The graph topology, i.e. the connections between the nodes,  is fully characterized by the adjacency matrix  $A$ of the graph, whose elements $A_{jk}$ correspond to  the edges:
\begin{equation}
\label{eq:adjacency_matrix}
A_{jk} = \begin{cases}
1 & \text{if  $j \neq k$ and $(j, k) \in \mathcal{E}$,}\\
0 & \text{otherwise.} 
\end{cases}
\end{equation}
The adjacenct matrix is a valid generator for the dynamics of CTQWs, and indeed a common choice for its Hamiltonian. On the hand, the quantum nature of the system allows us to consider also non-real but 
Hermitian matrices \cite{Lu2016,frigerio2021generalized,cavazzoni2022perturbed}, as
\begin{equation}
\label{eq:adjacency_matrix_chiral}
H_{jk} = A_{jk}\, e^{-i\phi_{jk}},
\end{equation}
where the phases $\phi_{jk}\in [0,2\pi)$ (satisfying $\phi_{jk}=-\phi_{kj}$ to ensure the Hermiticity condition) describes the effect of local gauge (e.g., magnetic) fields \cite{frigerio2021generalized}.
Weighting the edges of the network is also possible, and corresponds to tuning the moduli of the transition 
amplitudes in the Hamiltonian, e.g. setting $|H_{jk}|=\beta_{jk} \in \mathbb{R}_+$. 
In turn, the edge weights can be controlled, by changing the distance among the elements of the network \cite{peruzzo10,sciarrino12,crespi21,PhysRevB.76.115332,Roushan}.
\subsection{Qubit quantum routing by CTQW} 
An \textit{ideal} quantum router is a distributed system that takes an input quantum state supported by  a small subset of nodes  $\{\ket{x_j^{\rm{in}}}\}$   and deterministically  transfers it to a  designated target subset $\{\ket{x_j^{\rm{out}}}\}_{\xi}$,   without any losses or decoherence. The target location  is selected by tuning the Hamiltonian parameters $\xi$, such as couplings and phases. 
The time evolution associated with the transferred quantum state, setting $\hbar=1$, is given  by
\begin{equation}
    \label{eq:ideal quantum routing}
     \ket{\psi^{\rm{out}}}_{\xi} = U(\xi,t^*) \ket{\psi^{\rm{in}}},\qquad U(\xi,t^*)= e^{-i H(\xi)\,t^{*}},
\end{equation}
where 
$t^*$ is the routing time at which the  evolution achieves perfect state transfer to the specified output nodes.
The time evolution operator effectively acts as a projector from the initial state into the desired output state $\ket{\psi^{\rm{out}}}_{\xi} \bra{\psi^{\rm{in}}}$. For the routing a generic quantum state
\begin{equation}
     \ket{\psi^{\rm{in}}} = \sum_{j} \psi_{j} \ket{x^{\rm{in}}_{j}} \xrightarrow[]{\rm{routing} }\ket{\psi^{\rm{out}}}_{\xi} = \sum_{j} \psi_{j}  \ket{x^{\rm{out}}_{j}}_{\xi},
\end{equation}
where the amplitudes $\psi_j$ are preserved while the support of the state is changed from the input nodes  to the target ones.
In order for the routing scheme to work correctly, the dynamics must exhibit a universal routing time $t^{*}$, that does not depend on the input state $\ket{\psi^{\rm{in}}}$. Moreover, the input $\ket{\psi^{\rm{in}}}$ and all the possible output states $\{ \ket{\psi^{\rm{out}}}_{\xi} \}$ must be mutually orthogonal, i.e. $\braket{\psi^{\rm{in}}}{\psi^{\rm{out}}}_{\xi} = 0 \,\, \forall \xi$ and $_{\nu}\braket{\psi^{\rm{out}}}{\psi^{\rm{out}}}_{\xi} = 0 \, \forall \, \xi,\nu$ with $\xi \neq \nu$,  ensuring that each routed state is perfectly distinguishable from both the initial state and  from the states associated with other targets. Finally, the protocol must allow for the possibility to select the desired output through the controlled tuning of the Hamiltonian parameters, such that $\ket{\psi^{\rm{out}}}_{\xi} = U(\xi,t^*) \ket{\psi^{\rm{in}}}$ and $\ket{\psi^{\rm{out}}}_{\nu} = U(\nu,t^*) \ket{\psi^{\rm{in}}}$.
\newline
As the basic building block of quantum technologies is the qubit, we focus on the routing of qubit states. The state of the qubit is encoded in the position of the walker, i.e., in the superposition states over two nodes of a graph $\{\ket{x^{\rm{in}}_1}_{\xi},\ket{x^{\rm{in}}_2}_{\xi}\}$, such that the input and output qubit  take the form:
\begin{align}
    \label{eq:input_output_state}
    &\ket{\psi^{\rm{in}}(\alpha,\chi)} = \cos{\left(\frac{\alpha}2\right)} \ket{x^{\rm{in}}_1} + e^{i\chi}\sin{\left(\frac{\alpha}2\right)} \ket{x^{\rm{in}}_2} \nonumber \\
    &\ket{\psi^{\rm{out}}(\alpha,\chi)}_{\xi} = \cos{\left(\frac{\alpha}2\right)} \ket{x^{\rm{out}}_1}_{\xi} + e^{i\chi}\sin{\left(\frac{\alpha}2\right)} \ket{x^{\rm{out}}_2}_{\xi},
\end{align}
with $\alpha \in [0,\pi]$ and $\chi\in\left[0,2\pi\right)$ and the pair of nodes $\{\ket{x^{\rm{out}}_1}_{\xi},\ket{x^{\rm{out}}_2}_{\xi}\}$ identifies  the target location, which is selected through the choice of the Hamiltonian parameters 
$\xi$.

A topology enabling perfect quantum routing has been put forward in \cite{cavazzoni2025perfect} and 
termed \textit{Lily Graph}.  Its structure is depicted in Fig.\ref{fig:Lily Graph}. We also notice 
that other networks may support perfect quantum routing. Indeed, every graph which is reduced to the 
Lily Graph grouping together identically evolving vertices inherits the same time evolution properties, and consequently the operational function of quantum routing. 
The Lily router consists of a central set of nodes connected to several identical branches, arranged so that quantum states can be routed from any chosen input branch to any selected output branch with unit fidelity independently of the form of the initial wave-function. 
The aim is to route an initial qubit encoded in nodes $\ket{1}$ and $\ket{2}$ into the qubit identified with the vertices $\ket{r}$ and $\ket{f}$.
The  Hamiltonian corresponding to the Lily graph is the sum of the adjacency matrices of the different layers of the structure, i.e. the input layer $\mathcal{I}$, the chiral layer $\mathcal{C}$, the routing layer $\mathcal{R}$ and the output layer $\mathcal{O}$, as: 
\begin{align}
    \label{eq:Lily_Hamiltonian}
     H  \left( n,\beta,\phi_{0},\phi_{1},\phi_{2} \right) = A_\mathcal{I} + \beta C_{\mathcal{C}} \left( \phi_{0} \right) + \beta C_{\mathcal{R}} \left( n,\phi_{1},\phi_{2} \right) +A_\mathcal{O} ( n ).
\end{align}
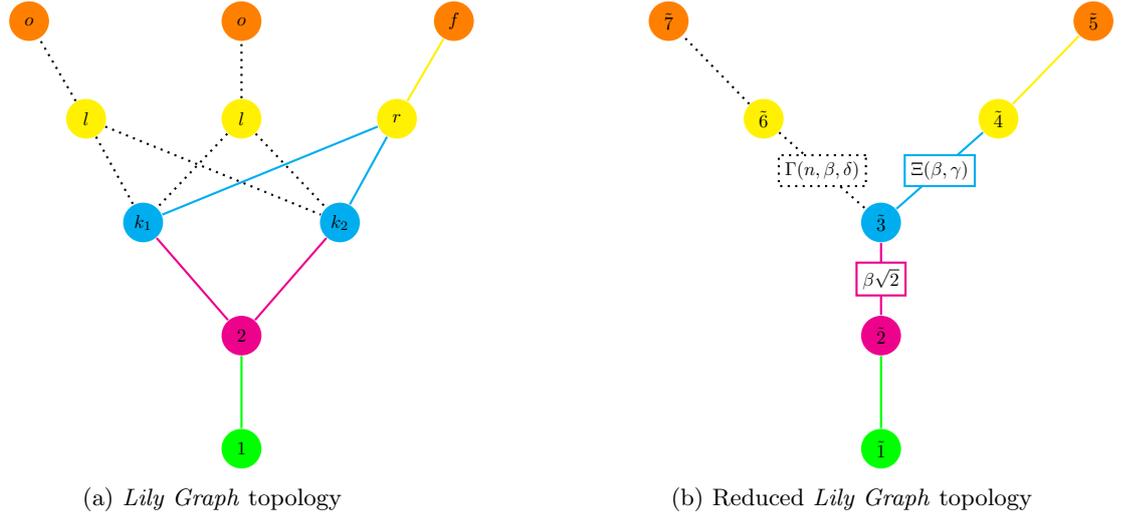
\begin{figure}[!h]
  \centering
  \begin{subfigure}[b]{0.45\textwidth}
    \raggedleft
    \begin{tikzpicture}[thick,scale=0.7, every node/.style={scale=0.7}]
    \tikzset{root node/.style={circle,fill=green,minimum size=0.75cm,inner sep=0pt}}
    \tikzset{two node/.style={circle,fill=magenta,minimum size=0.75cm,inner sep=0pt}}
    \tikzset{chiral node/.style={circle,fill=cyan,minimum size=0.75cm,inner sep=0pt}}
    \tikzset{constructive node/.style={circle,fill=yellow,minimum size=0.75cm,inner sep=0pt}}
    \tikzset{f node/.style={circle,fill=orange,minimum size=0.75cm,inner sep=0pt}}
    \tikzset{ghost node/.style={circle,fill=white,minimum size=0.75cm,inner sep=0pt}}
    
    \tikzset{dots node/.style={circle,fill=white!40,minimum size=0.75cm,inner sep=0pt}}
      \node[f node] (5) {$o$};
      \node[constructive node] (4) [below = 0.75cm of 5]  {$l$};
      \node[ghost node] (3) [below = 2.125cm of 5]  {};
      \node[two node] (2) [below = 3.625cm of 5]  {$2$};
      \node[root node] (1) [below = 5.125cm of 5]  {$1$};
      \node[chiral node] (6) [left = 0.75cm of 3]  {$k_1$};
      \node[chiral node] (8) [right = 0.75cm of 3]  {$k_2$};
      \node[constructive node] (10) [left = 1.5cm of 4]  {$l$};     
      \node[constructive node] (11) [right = 1.5cm of 4]  {$r$}; 
      \node[f node] (12) [left = 2.25cm of 5]  {$o$};     
      \node[f node] (13) [right = 2.25cm of 5]  {$f$}; 
      \path[draw=green,thick]
      (1) edge node {} (2);
      \path[draw=magenta,thick]
      (2) edge node {} (6)
      (2) edge node {} (8);
      \path[draw=cyan,thick]
      (11) edge node {} (6)
      (11) edge node {} (8);
      \path[draw=yellow,thick]
      (11) edge node {} (13);
      \path[draw,dotted]
      (10) edge node {} (6)
      (10) edge node {} (8);
      \path[draw,dotted]
      (4) edge node {} (6)
      (4) edge node {} (8);
      \path[draw,dotted]
      (4) edge node {} (5)      
      (10) edge node {} (12);
    \end{tikzpicture}
    \caption{\textit{Lily Graph} topology}
    \label{fig:Lily Graph}
  \end{subfigure}
  \hfill
  \begin{subfigure}[b]{0.45\textwidth}
    \raggedleft
    \begin{tikzpicture}[thick,scale=0.4, every node/.style={scale=0.7}]
    \tikzset{root node/.style={circle,fill=green,minimum size=0.75cm,inner sep=0pt}}
    \tikzset{two node/.style={circle,fill=magenta,minimum size=0.75cm,inner sep=0pt}}
    \tikzset{chiral node/.style={circle,fill=cyan,minimum size=0.75cm,inner sep=0pt}}
    \tikzset{constructive node/.style={circle,fill=yellow,minimum size=0.75cm,inner sep=0pt}}
    \tikzset{f node/.style={circle,fill=orange,minimum size=0.75cm,inner sep=0pt}}
    \tikzset{ghost node/.style={circle,fill=white,minimum size=0.75cm,inner sep=0pt}}
    
    \tikzset{dots node/.style={circle,fill=white!40,minimum size=0.75cm,inner sep=0pt}}
      \node[ghost node] (5) {};
      \node[ghost node] (4) [below = 0.75cm of 5]  {};
      \node[chiral node] (3) [below = 2.125cm of 5]  {$\tilde 3$};
      \node[two node] (2) [below = 3.625cm of 5]  {$\tilde 2$};
      \node[root node] (1) [below = 5.125cm of 5]  {$\tilde 1$};
      \node[ghost node] (6) [left = 0.75cm of 3]  {};
      \node[ghost node] (8) [right = 0.75cm of 3]  {};
      \node[constructive node] (10) [left = 1.0cm of 4]  {$\tilde 6$};     
      \node[constructive node] (11) [right = 1.0cm of 4]  {$\tilde 4$}; 
      \node[f node] (12) [left = 2.25cm of 5]  {$\tilde 7$};     
      \node[f node] (13) [right = 2.25cm of 5]  {$\tilde 5$}; 
      \path[draw=green,thick]
      (1) edge node {} (2);
      \path[draw=magenta,thick]
      (2) edge node[rectangle, fill=white, draw] {$\beta\sqrt{2}$} (3);
      \path[draw=cyan,thick] 
      (11) edge node[rectangle, fill=white, draw] {$\Xi(\beta,\gamma)$} (3);
      \path[draw=yellow,thick]
      (11) edge node {} (13);
      \path[draw,double,dotted]
      (10) edge node[rectangle, fill=white, draw] {$\Gamma(n,\beta,\delta)$} (3);
      \path[draw,double,dotted]
      (10) edge node {} (12);
    \end{tikzpicture}
    \caption{Reduced \textit{Lily Graph} topology}
    \label{fig:Reduced Lily Graph}
  \end{subfigure}
  \caption{(a)-- The {input nodes} ($\ket{1}$ and $\ket{2}$) are green and magenta labelled, the \textit{chiral layer} ($\{\ket{k_j}\}$), composed of $2$ nodes, is blue labelled, the \textit{routing layer} ($\ket{l}$ and $\ket{r}$) composed of $n$ vertices is yellow labelled and the $n$ outputs ($\ket{f}$ and $\ket{o}$) are orange labelled. Here for simplicity the structure present 3 outputs, but in general the number of outputs can be arbitrary high, as theoretically presented in Eq.\eqref{eq:out_adjacency}. (b)-- Reduced Lily Network. Exploiting the symmetry of the structure, it is possible to define a reduced network, in which the vertices associated with the same quantum probability are grouped together. The colour palette reproduce the one presented in Fig.\ref{fig:Lily Graph}. The mathematical definition of each node is presented in Eq.\eqref{eq:Symmetry Basis}.}
  \label{fig:Lily}
\end{figure}
$A_\mathcal{I}$ refers to the input adjacency matrix, i.e.
\begin{equation}
    \label{eq:input_adjacency}
    A_\mathcal{I} = \ket{1}\bra{2} + \ket{2}\bra{1},
\end{equation}
and is associated to the {input} state $\ket{\psi^{\rm{in}}}$. Then, there is the chiral adjacency $C_{\mathcal{C}}$ contribution which reads:
\begin{equation}
    \label{eq:deph_adjacency}
    C_{\mathcal{C}}\left( \phi_0 \right) = e^{i\phi_{0}}\ket{2}\bra{k_1} +\ket{2}\bra{k_2} + h.c. \, ;
\end{equation}
the routing adjacency $C_{\mathcal{R}}( n, \phi_1, \phi_2)$, defined as
\begin{align}
    \label{eq:int_adjacency}
    C_{\mathcal{R}} \left( n, \phi_1, \phi_2 \right) = \sum_{\substack{l \in \mathcal{R} \\ l\neq r}}^{n} \bigg( e^{i\phi_{1}}\ket{k_1} \bra{l} + \ket{k_2}\bra{l} \bigg) + e^{i\phi_{2}} \ket{r}\bra{k_1} + \ket{r}\bra{k_2} + h.c. \, ,
\end{align}
and finally, the output adjacency $A_\mathcal{O}$ which is given by
\begin{align}
    \label{eq:out_adjacency}
    A_\mathcal{O} \left( n \right) = \ket{r} \bra{f} + \ket{f} \bra{r} + \sum_{\substack{l \in \mathcal{R} \\ l \neq r}}^{n} \sum_{\substack{o \in \mathcal{O} \\ o \neq f}}^{n} \left( \ket{l}\bra{o} + \ket{o} \bra{l} \right).
\end{align}
Since in principle the number of outputs can be arbitrarily high, to efficiently study the quantum dynamics over the \textit{Lily graph} it is necessary to resort to dimensionality reduction techniques \cite{novo2015systematic,chakraborty2016spatial}. Grouping identically evolving vertices \cite{meyer2015connectivity}, it is possible to get an orthonormal basis for the Lily Graph. The new basis, which preserve the properties of dynamics in the graph, but with a lower Hilbert space dimension, reads  
\begin{align}
    \label{eq:Symmetry Basis}
    \ket{\tilde 1}=& \ket{1} \,, \quad
    \ket{\tilde 2}= \ket{2} \,, \nonumber \\
    \ket{\tilde 3}=& \frac{1}{\sqrt{2}}\left( e^{-i\phi_{0}} \ket{k_1} + \ket{k_2} \right) \nonumber \\
    \ket{\tilde 4}=& \ket{r}\,, \quad
    \ket{\tilde 5}= \ket{f} \nonumber \\
    \ket{\tilde 6}=& \frac{1}{\sqrt{n-1}}\sum_{\substack{l \in \mathcal{R} \\ l \neq r}}^{n} \ket{l} \nonumber \\ 
    \ket{\tilde 7}=& \frac{1}{\sqrt{n-1}} \sum_{\substack{o \in \mathcal{O} \\ o \neq f}}^{n} \ket{o} \nonumber \\
\end{align}
In such basis, it is possible to write a reduced Hamiltonian that reproduces the quantum dynamics of the overall structure, as $\tilde H_{i,j}=\bra{\tilde i}H\ket{\tilde j}$
\begin{equation}
    \label{eq:reduced_Hamiltonian_beta}
    \tilde H (n,\beta,\gamma, \delta) =
    \begin{pmatrix}
 	0 & 1 & 0 & 0 & 0 & 0 & 0 \\
	1 & 0 & \beta \sqrt{2} & 0 & 0 & 0 & 0 \\
	0 & \beta \sqrt{2} & 0 & \Xi(\beta,\gamma) & 0 & \Gamma(n,\beta,\delta) & 0 \\
	0 & 0 & \Xi^{*}(\beta,\gamma) & 0 & 1 & 0 & 0 \\
    0 & 0 & 0 & 1 & 0 & 0 & 0 \\
    0 & 0 & \Gamma^{*}(n,\beta,\delta) & 0 & 0 & 0 & 1 \\
    0 & 0 & 0 & 0 & 0 & 1 & 0 \\
\end{pmatrix} ,
\end{equation}
with the coefficients $\Xi(\beta,\gamma) = \beta (1+e^{-i(\phi_{1}+\phi_{0})})/\sqrt{2}$ and $\Gamma(n,\beta,\delta) = \beta \sqrt{(n-1)/2}(1+e^{-i(\phi_{2}+\phi_{0})})$ depending only on  the effective phases $\gamma=-(\phi_{1}+\phi_{0})$ and $\delta=-(\phi_{2}+\phi_{0})$. These two coefficients govern two distinct properties of the structure. $\Xi(\beta,\gamma)$ is related to the transition probability between the input and output state, while $\Gamma(n,\beta,\delta)$ is related to the probability of unwanted routing, as it mathematically represent the connection of the undesired output with the rest of the network (see Fig.\ref{fig:Reduced Lily Graph}). Upon setting the value of $\beta=\sqrt{3}/2$ and $\phi_{0}=\phi_{1}=\pi \, , \phi_{2}=0$ at time $t^{*}=\pi+2m\pi$, with $m \in \mathbb{N}$, the time evolution operator is reduced to \cite{cavazzoni2025perfect}
\begin{align} 
    \label{eq: Projector}
    &\tilde U(t^{*}=\pi+2m\pi) = e^{-i\tilde Ht^{*}}    
    = \ket{\tilde 1}\bra{\tilde 5} + \ket{\tilde 2}\bra{\tilde 4} + \ket{\tilde 3}\bra{\tilde 3} + \ket{\tilde 5}\bra{\tilde 1} + \ket{\tilde 4}\bra{\tilde 2} + \ket{\tilde 6}\bra{\tilde 7} + \ket{\tilde 7}\bra{\tilde 6},
\end{align} 
and then the routing procedure perfectly transport the initial qubit 
\begin{equation} \label{eq:routing_qubits}
     \ket{\psi^{\rm{in}}(\alpha,\chi)}=\cos{\left(\frac{\alpha}2\right)} \ket{\tilde 1}+e^{i\chi}\sin{\left(\frac{\alpha}2\right)}\ket{\tilde 2} 
\end{equation}
into the target qubit
\begin{equation}
    \ket{\psi^{\rm{out}}(\alpha,\chi)}_{\xi}=\cos{\left(\frac{\alpha}2\right)} \ket{ \tilde{5}}+e^{i\chi}\sin{\left(\frac{\alpha}2\right)}\ket{ \tilde{4}}.
\end{equation}
Summarizing: the structure supports ideal quantum routing from the initial nodes nodes $\ket{\tilde 1}=\ket{1}$ and $\ket{\tilde 2}=\ket{2}$ to the desired output nodes $\ket{\tilde 4}=\ket{r}$ and $\ket{\tilde 5}=\ket{f}$, excluding from the dynamics all the unwanted output regions. The position of the phases allows the selection among the possible outputs, acting as the parameter $\xi$ that discerns the $n$ outputs. Additionally, as expressed by Eq.\eqref{eq: Projector}, the symmetry of the structure guarantees that each selected output can also act as an input and, vice-versa, the input can act as an output region. This property guarantees that the Lily Graph is a network that support quantum information exchange among $n$ senders and receivers (see Appendix \ref{app:Comparison} for further details). 

On the other hand, in physical implementations the fine tuning of the parameters $\beta$, $\gamma$ and $\delta$ is inevitably  affected by some uncertainty due to noise sources, leading to non-perfect routing protocols. 
In the following we will analyse the impact of different noise models on the routing procedure to assess its robustness  against  static imperfections and dynamical fluctuations in the parameters $\beta$, $\gamma$ and $\delta$.
\subsection{Phase Noise Models}
The routing procedure strongly depends on the value of the phases $\gamma$ and $\delta$ in Eq.\eqref{eq:reduced_Hamiltonian_beta}. If $\gamma=\pi$ the coefficient $\Gamma(n,\beta,\delta)=\Gamma(\beta,\gamma)=0$ and the dynamics of the systems involves only the desired output region among all the possible $n$ outputs. Contrariwise, if for some noise effect $\gamma \neq \pi$ the information may also be routed to the undesired outputs, leading to potential loss of information. Additionally, the coefficient $\Xi\left(\beta,\delta\right)$ is a source of noise  if $\delta\neq 0$, as it changes the transition amplitude 
from the input to the output spatial region. 
\subsubsection{Static Phase Noise}
As a first noise model, we analyse the effects of static fluctuations in the phases $\gamma$ and $\delta$, as $\gamma_{\rm{noisy}} = \gamma + \epsilon_1$ and $\delta_{\rm{noisy}} = \delta + \epsilon_2$, in which the two independent random variables $\{\epsilon_1,\epsilon_2\}$ follow the von Mises distribution \cite{sra2012short,kurz2016kullback,ragazzi2024generalized}:
\begin{equation}   
    p_k(\epsilon)=\frac{e^{k\cos{\epsilon}}}{2\pi I_0(k)}
\end{equation}
where $I_0(k)$ is the modified Bessel function of the first kind. The parameter $k$ is a concentration parameter, which controls  the distribution's width: larger  $k$ lead to a tighter distribution around $\epsilon=0$. We assume that the $k$ is the same for the distribution width of both $\epsilon_{1}$ and $\epsilon_{2}$. 
This working condition is physically justified by the experimental procedure that induce the phases, which consist in the application of local magnetic fields \cite{yasser17}. 
For small values of $k$, the von Mises distribution is well approximated by a Gaussian with variance 
$\sigma^2=1/k$.
Starting from an initial state $\vert \psi^{\rm{in}}\rangle$, the evolved state in the presence of 
static phase noise be expressed as:
\begin{align}
    \label{eq:sigma_out}
    &\rho(\alpha,\chi,n,t,\beta,\gamma,\delta,k)= \\
    &\int_{-\pi}^{\pi} \int_{-\pi}^{\pi} d\epsilon_{1} \,d\epsilon_{2} \,p_k(\epsilon_{1},\epsilon_{2})\tilde U(n,t,\beta,\gamma+\epsilon_{1},\delta+\epsilon_{2})\vert \psi^{\rm{in}} (\alpha,\chi) \rangle\langle \psi^{\rm{in}} (\alpha,\chi) \vert \tilde{U}^{\dagger}(n,t,\beta,\gamma+\epsilon_{1},\delta+\epsilon_{2}), \nonumber
\end{align}
where the single realization evolution operator is $\tilde U(n,t,\gamma+\epsilon_{1},\delta+\epsilon_{2})=e^{-i \tilde H(n,t,\gamma+\epsilon_{1},\delta+\epsilon_{2}) t}$. 
\subsubsection{Dynamical Phase Noise}
To include dynamical noise in the routing scheme, we model  phase fluctuations using a Ornstein–Uhlenbeck (OU) processes \cite{maller2009ornstein,ragazzi2025scalable}. OU noise is a  Gaussian, mean-reverting, stochastic process  $\{X_t\}_t$  governed by the  differential equation:
\begin{align}
dX_{t}=\theta(\mu-X_t)dt +\Sigma\,dW_t
\label{eq:oup}
\end{align}
where $\mu$ is the long-time mean of the process, $\theta$ is the mean reversion speed, $\Sigma$ the volatility and $dW_t$ is a Wiener increment. 
For a random initial condition, the variable $X_{t}$ follows a normal distribution with variance $\sigma^2=\frac{\Sigma^2}{2\theta}$ for all $t$, including  the initial time. 
For small $k$, the OU process may be seen as the dynamical counterpart of the von Mises noise through the approximation $\sigma^2 \simeq 1/k$. 

If OU noise affects the chiral phases, the Hamiltonian becomes time-dependent and the dynamics of an initial state $\rho_0=\ket{\psi^{\rm{in}}} \bra{\psi^{\rm{in}}}$ over the router is described as an ensemble average over all realizations of the stochastic process, that is, 
\begin{align}
{\rho}(\alpha,\chi,n,t,\beta,\gamma,\delta,\Sigma,\theta)=\left\langle \tilde U(n,t,\beta,\gamma,\delta,\Sigma,\theta)\,\rho_0 (\alpha,\chi) \,\tilde U^{\dagger}(n,t,\beta,\gamma,\delta,\Sigma,\theta) \right\rangle_{X}
\end{align}
where $\tilde U(n,t,\beta,\gamma,\delta)=\mathcal{T} e^{-i\int_0^tdt' \tilde H(n,\beta, \gamma(t^{'}),\delta(t^{'}))}$  is the evolution operator generated by the time-dependent Hamiltonian $\tilde H(n, \beta, \gamma(t),\delta(t))$, $\mathcal{T}$ is the time ordering operator,  $\gamma(t) $ and $\delta(t)$ are stochastic phases describing a OU process according to Eq.\eqref{eq:oup} and $\langle\cdot\rangle_X$ denotes the ensemble average over the realizations of the noise process $X$.
\subsection{Weight Noise Models}
Fluctuations in the coupling strengths between the nodes of the network  \cite{zwick2011robustness} also affect the routing performance. Previous works have shown  that an appropriate distribution of couplings weight  in spin network is fundamental for achieving  high fidelity state transfer \cite{bose2003quantum,burgarth2005perfect,burgarth2005conclusive}, and the same  applies to quantum 
routing \cite{cavazzoni2025perfect}. As for phase noise, we examine the influence of both static and dyanmical
noise. Differently from the phase noise models, however, a sub-optimal value of the parameter $\beta$ 
does not route the quantum information to unintended outputs. Instead, it only changes the probability of finding the quantum state in the chosen output.  This difference arises because the phases determine which outputs participate in the dynamics of the initial state. Indeed, they induce destructive interference in all undesired regions while enabling 
 constructive interference only at the selected output. The weight $\beta$, on the other hand, ensures a unit transition probability from the input to the output region \cite{cavazzoni2025perfect}.
\subsubsection{Static Weight Noise}
Static fluctuations in the weight $\beta$ may be described by writing 
$\beta_{noisy} = \beta + \zeta$, where $\zeta$ is drawn from a Gaussian distribution: 
\begin{equation}
    \label{eq:Gaussian}
    p_{\sigma}(\zeta)=\frac{e^{-\frac{\zeta^2}{2\sigma^2}}}{\sigma\sqrt{2\pi}}.
\end{equation}
Unlike the  phase parameters, here we employ a Gaussian distribution because $\beta$ is not restricted  to a compact domain, in contrast to $\gamma$ and $\delta$. Nonetheless, this Gaussian distribution is comparable with the von Mises one, through the above-mentioned correspondence $\sigma^2 \rightarrow 1/k$, valid for small $k$. As for the phase noise, the time evolution of the initial state $\vert \psi^{\rm{in}}\rangle$ in such noisy network can be expressed as 
\begin{align}
    &\rho(\alpha,\chi,n,t,\beta,\gamma,\delta,\sigma)= \int_{-\infty}^{\infty} d\zeta \,p(\zeta)\tilde U(n,t,\beta,\gamma,\delta,\zeta)\ket{\psi^{\rm{in}} (\alpha,\chi) } \bra{\psi^{\rm{in}} (\alpha,\chi) } \tilde{U}^{\dagger}(n,t,\beta,\gamma,\delta,\zeta).
\end{align}
\subsubsection{Dynamical Weight Noise}
The dynamic weight noise is modelled using an OU process, following the same framework used for the phase parameters in Eq.~\eqref{eq:oup}. In this case, however, the stochastic perturbation acts on the coupling weight $\beta$, rather than on the chiral phases of the Hamiltonian.
\section{Results}
\label{sec:R}
A central step in assessing the performance of a quantum routing protocol is to quantify how different noise mechanisms affect its ability to faithfully transfer quantum information.
Given a specific initial state, the routing fidelity quantifies the overlap between the time evolved state and the desired target output state.
Since no preferred input  state  undergoes the routing procedure, the performances of the Lily network is characterized by evaluating the  average routing fidelity in the presence of noise.  This quantity is obtained by averaging the fidelity over all possible input states, as defined in Eq. \eqref{eq:routing_qubits}, on the Bloch sphere. 
The average routing fidelity in the presence of noise, may thus written as 
\begin{equation}
    \bar{ F}(n,t, \vec{\eta}_{H}^{\, \rm{opt}}, \vec{\eta}_{n})=
\frac1{4\pi}\int_0^{\pi}\!\!\!\sin\alpha\,d\alpha\int_0^{2\pi}\!\!\!d\chi \,\bra{\psi^{\rm{out}}(\alpha,\chi)}\rho(\alpha, \chi,n,t, \vec{\eta}_{H}^{\, \rm{opt}}, \vec{\eta}_{n})\ket{\psi^{\rm{out}}(\alpha,\chi)},
\end{equation} 
where, for simplicity, we omitted the  subscript that specifies the dependence of the target output state on the protocol parameters. 
The optimal set of  Hamiltonian parameters (i.e, $\beta=\frac{\sqrt3}2, \gamma=0$ and $\delta=\pi$) is denoted by  $\vec{\eta}_{H}^{\, \rm{opt}} $, while $ \vec{\eta}_{n} $  refers to the parameters associated with the chosen noise model. Depending on the kind of noise, $ \vec{\eta}_{n} $ may represent  the concentration parameter $k$ for the von Mises distribution, the drift $ \theta$ or the volatility $\Sigma$ for the OU noise or to the variance $\sigma^2$ of a Gaussian distribution. 
Any reduction in the mean routing fidelity compared to the noiseless case is a signature of imperfect routing process.  The detailed derivation of the fidelity expressions for each noise source is reported in the Appendix \ref{app:noise}. For notational convenience, in the following we will refer to the mean fidelity simply as $\bar{F}$, keeping its  dependence on all the parameters $(n,t, \vec{\eta}_{H}^{\, \rm{opt}}, \vec{\eta}_{n})$ implicit.
\subsection{Phase Fluctuations}
In this section, we discuss how static and dynamical fluctuations of the 
phases $\delta$ and $\gamma$ impact the routing protocol.  In both the static and dynamic scenarios,   
the analysis for different number of outputs is required, since the noisy phases no longer guarantee perfect constructive interference at the desired output nor destructive interference at the unwanted ones.
For this reason, we consider routing networks with  $n=2,10,30$ outputs and, for each value of $n$, we evaluate the effect of both static and dynamic phase noise. 
\subsubsection{Static Phase Noise}
\begin{figure}[t]
  \begin{subfigure}[b]{0.5\textwidth}
    \centering
    \includegraphics[width=\linewidth]{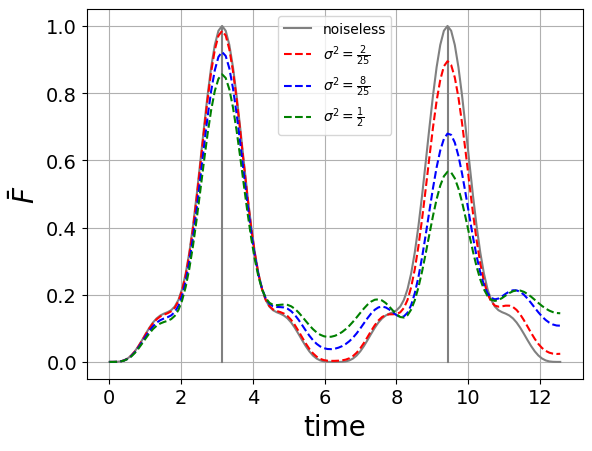}
    \caption{Static Phase Noise Mean Fidelity}
    \label{fig:static phase noise mean fidelity}
  \end{subfigure}
  \begin{subfigure}[b]{0.5\textwidth}
    \centering
    \includegraphics[width=\linewidth]{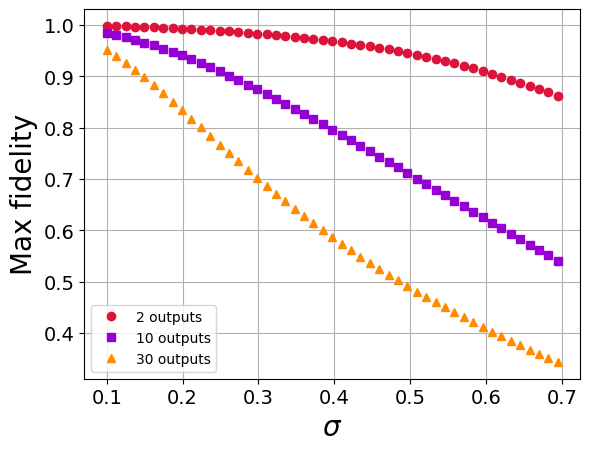}
    \caption{ Static Phase Noise Maximum Fidelity Scaling}
    \label{fig:Robustness Fidelity vM}
  \end{subfigure}
  \caption{\textbf{Effect of static Phase Noise} -- (a) Average quantum fidelity as a function of time, for different values of the concentration parameter $k$ and for $n=2$ outputs. To facilitate the comparison with the dynamical noise, we label the curves with the effective Gaussian variance $\sigma^2=\frac1k$. The continuous gray curve is the unperturbed unitary case and the vertical grey lines mark its first peaks, located at $t=\pi$ and $t=3\pi$.  (b) Scaling of the maximum of the average fidelity as a function of the effective Gaussian standard deviation $\sigma=\sqrt{\frac1k}$, for different number of outputs.}
  \label{fig:von Mises phase noise and fidelity scaling}
\end{figure}

In the presence of static phase noise, the main effect of the noise is a progressive degradation of the periodic ideal routing of the initial state towards the target output. 
As shown in Fig.\ref{fig:static phase noise mean fidelity}, an increase in  the noise strength leads to a gradual reduction of the  height of the fidelity peaks\footnote{The noise strength  is represented by the parameter $\sigma$ in Fig.\ref{fig:von Mises phase noise and fidelity scaling}. This parameter is related to the von Mises concentration parameter $k$ through the correspondence  $\sigma^2 \approx 1/k$. We use the quantity $\sigma$ here because it allows for a more direct   comparison with the other noise models considered in this work}. 
This behaviour occurs because phase fluctuations  perturb the value of the parameter $\Xi(\beta,\delta)$, hence modifying the transition probability between the input and the selected output region. However, the protocol is quite robust at the first ideal routing time $t^{*}=\pi$, which still exhibits  a considerably high fidelity peak within the considered noise range. 
As noted above, the constructive and destructive interference condition depends on the number of outputs $n$, which in turn determines the robustness of the overall procedure. In Fig.\ref{fig:Robustness Fidelity vM} different shapes and colours correspond to a different number of outputs $n$, showing that  larger number $n$ of outputs results in a more significant impact of  noise on the routing process. This  increased sensitivity arises because a higher number of outputs modifies the parameter $\Gamma(n,\beta,\delta)$ appearing in Eq.~\eqref{eq:reduced_Hamiltonian_beta}, enhancing unwanted transmission toward non-selected outputs.


\subsubsection{Dynamical Phase Noise} 
\begin{figure}[ht!]
  \begin{subfigure}[b]{0.5\textwidth}
    \centering
    \includegraphics[width=\linewidth]{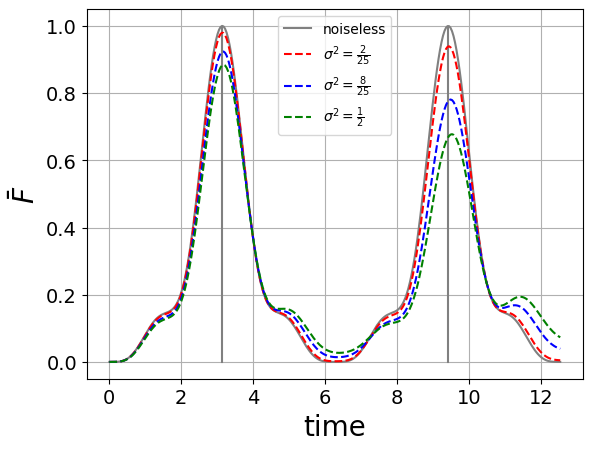}
    \caption{OU Phase Noise Mean Fidelity}
    \label{fig:dynamic noise mean fidelity}
  \end{subfigure}
  \begin{subfigure}[b]{0.5\textwidth}
    \centering
    \includegraphics[width=\linewidth]{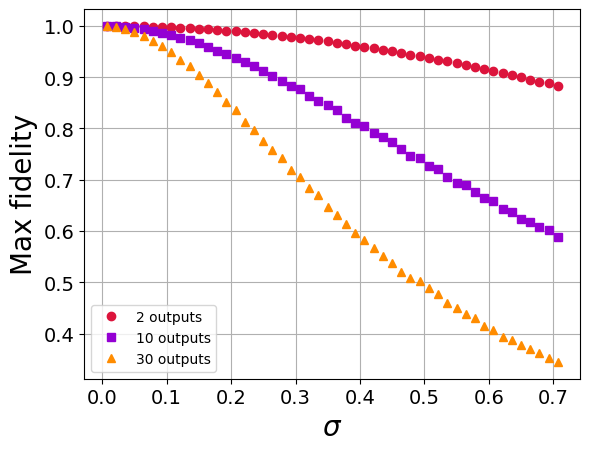}
    \caption{OU Phase Noise Maximum Fidelity Scaling}
    \label{fig:Robustness Fidelity O.U.}
  \end{subfigure}
  \caption{\textbf{Effect of the OU dynamic Phase Noise}-- (a) Average quantum fidelity as a function of time, for $n=2$ outputs and for different values of the noise parameters $\theta$ and $\Sigma$. The parameter $\theta$ is kept fixed at the value of $1$ while  $\Sigma=0.4$ (green),  $\Sigma=0.8$ (blue) and $\Sigma=1$ (red). The curves are labelled with the effective Gaussian variance ($\sigma^2=\frac{\Sigma^2}{2\theta}$). The continuous gray curve is the unperturbed unitary case and the vertical grey lines mark its first peaks, located at $t=\pi$ and $t=3\pi$. (b) Scaling of the maximum of the average fidelity as a function of the standard deviation $\sigma=\sqrt{\frac{\Sigma^2}{2\theta}}$ of the effective Gaussian distribution for different numbers of outputs.}
  \label{fig:OU phase noise and fidelity scaling}
\end{figure}

We next examine the effects of time-dependent phase noise, modelled as an OU process. 
The dynamical noise model extends the static case by introducing temporal fluctuations of the phases, with the static scenario recovered  in the limit of infinitely slow dynamics. 
In this sense, the OU model provides a broader characterization of phase noise in routing protocols.
The behaviour observed under OU noise is  similar to the von Mises one. As shown in  Fig.
\ref{fig:Robustness Fidelity vM} and Fig.\ref{fig:Robustness Fidelity O.U.},  the scaling of the main fidelity peak is comparable. 
This trend is further confirmed by the time evolution of the average routing fidelity shown in Fig.\ref{fig:dynamic noise mean fidelity}. Consistently with the static scenario, the first fidelity peak remains  robust also against  dynamic noise. With the noise strength considered, the first fidelity peak occurs, also in this case, around $t^{*}=\pi$. This demonstrate that the routing behaviour is robust also under time-dependent phase  fluctuations.  
\subsection{Weight Fluctuations} We now proceed by examining the effect of the edge weight fluctuations. Even before performing the numerical analysis of the fluctuations in $\beta$, it is possible to identify a substantial difference between this noise and the phase fluctuation models. When the chiral phases are set to their ideal value, the weight  noise influences the fidelity of the routing protocol independently of the number of outputs $n$. This happens because the  chiral phases determine the routing destination, while the parameter $\beta$ only governs the magnitude of the transition probability between the input and the selected output state. Mathematically, this is reflected in the behaviour of the  parameter $\Gamma(\beta,\delta)$, which quantifies the  transmission of the information towards the non-selected outputs. If  the optimal chiral phases are chosen, then $\Gamma(\beta,\delta=0) = 0$ $\forall \, \beta$, ensuring that, regardless of the value of the weight parameter,  transmission from the input to any unwanted output remains zero, as long as the phases are set correctly.

\subsubsection{Static Weight Noise} 
\begin{figure}[!ht]
  \begin{subfigure}[b]{0.5\textwidth}
    \centering
    \includegraphics[width=\linewidth]{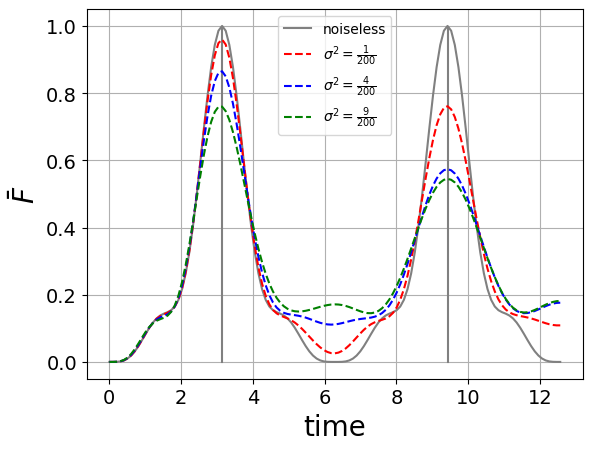}
    \caption{Gaussian Weight Noise Mean Fidelity}
    \label{fig:rumorebeta_n2}
  \end{subfigure}
  \begin{subfigure}[b]{0.5\textwidth}
    \centering
    \includegraphics[width=\linewidth]{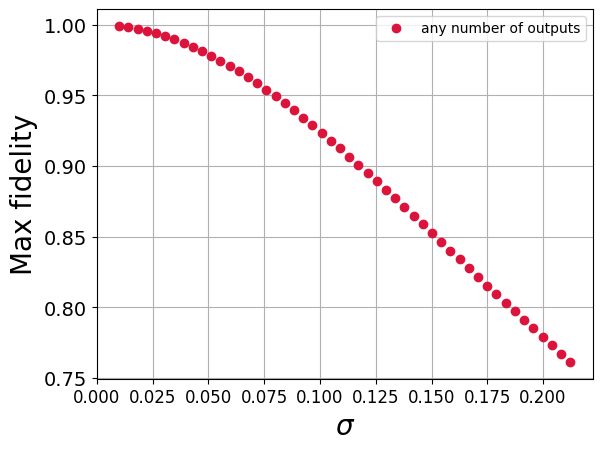}
    \caption{Gaussian Weight Noise Maximum Fidelity Scaling }
    \label{fig:Max_vs_sigma}
  \end{subfigure}
  \caption{\textbf{ Effect of static noise on the weights} --(a) Fidelity over time averaged with respect to the initial state, for  different values of the variance $\sigma^2$. The continuous gray curve is the unperturbed unitary case and the vertical grey lines mark its first peaks, located at $t=\pi$ and $t=3\pi$.\\
  (b) Reduction in the peak Fidelity as the standard deviation of the weight Gaussian distribution is raised.  }
  \label{fig:von Mises phases}
\end{figure}

The main effect of the static weight fluctuation consist in the decrease of the maximum fidelity as the noise strength increases. Nonetheless, as for the phase noise, the routing procedure proves quite robust at the first ideal routing time, i.e. $t^{*}=\pi$, showing a considerably high fidelity peak in the considered noise range, see Fig.\ref{fig:rumorebeta_n2}. 
Even if the effect of the noise is comparable to the phase fluctuation one, their origin is substantially different. The reduction of the transition probability from the input to the desired output state is originated from the explicit dependence of the Hamiltonian terms $\bra{\tilde 1} \tilde H \ket{\tilde 3} = \beta \sqrt{2}$ and $ \bra{\tilde 1} \tilde H \ket{\tilde 4} = \Xi(\beta,\gamma)$ over the weight parameter. A deviation from the ideal value of $\beta=\sqrt{3}/2$ directly influences the time evolution operator leading to a non perfect routing of information. 
In Fig.\ref{fig:Max_vs_sigma} we show how the first fidelity peak scales as a function of the noise strength (indepndently of the number of outputs $n$). 
\textcolor{black}{A direct comparison of these results with the phase fluctuation scaling in Fig.\ref{fig:Robustness Fidelity vM} shows that, even if weight fluctuations do not induce leakage of information towards undesired outputs, they degrade the height of the first fidelity peak faster than the phase noise for $n=2,10,30$ for a fixed value of $\sigma$. This indicates that for a moderate number of outputs (our numerical results have been obtained for $n \lesssim 30$), weight fluctuations constitute the main source of noise affecting routing performances.}

\subsubsection{Dynamic Weight Noise} 
\begin{figure}[!ht]
  \begin{subfigure}[b]{0.5\textwidth}
    \centering
    \includegraphics[width=\linewidth]{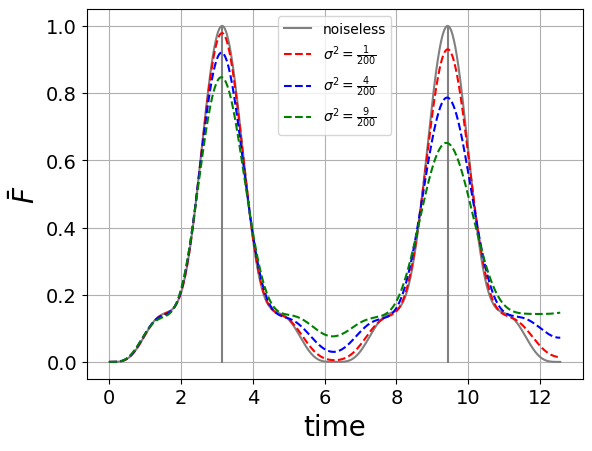}
    \caption{OU Weight Noise Mean Fidelity}
    \label{fig:Gaussian Weight}
  \end{subfigure}
  \begin{subfigure}[b]{0.5\textwidth}
    \centering
    \includegraphics[width=\linewidth]{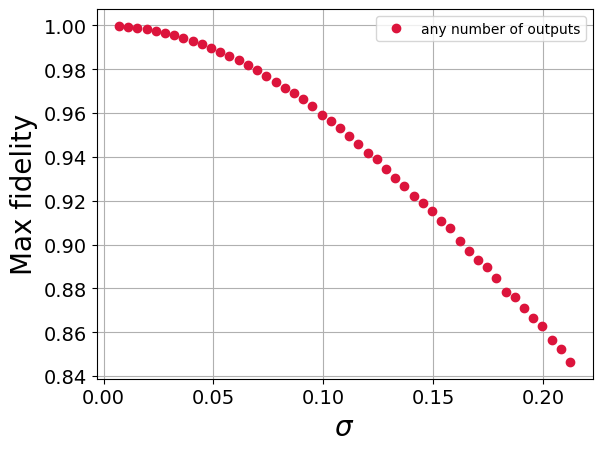}
    \caption{OU Weight Noise Maximum Fidelity Scaling }
    \label{fig:OU Weight Scaling}
  \end{subfigure}
  \caption{\textbf{Effect of dynamical noise on the weights} --(a) Average routing fidelity as a function of time, for different values of the noise parameters $\theta$ and $\Sigma$. The parameter $\theta$ is kept fixed at the value of $1$ while  $\Sigma=0.1$ (green),  $\Sigma=0.2$ (blue) and $\Sigma=0.3$ (red). The curves are labelled with the effective Gaussian variance ($\sigma^2=\frac{\Sigma^2}{2\theta}$). The continuous gray curve is the unperturbed unitary case and the vertical grey lines mark its first peaks, located at $t=\pi$ and $t=3\pi$. (b) Scaling of the maximum of the fidelity as a function of the standard deviation $\sigma=\sqrt{\frac{\Sigma^2}{2\theta}}$ of the effective Gaussian distribution.}
  \label{fig:OU Weight}
\end{figure}

Finally, we consider dynamical fluctuations of the weight $\beta$.
As in the case of dynamic phase noise, the OU model for weight fluctuations generalizes the static scenario.  
The qualitative agreement between Fig.\ref{fig:Max_vs_sigma} and Fig.\ref{fig:OU Weight Scaling} confirms it. The scaling of the 
 main fidelity peak exhibits the same qualitative behaviour in both static and dynamic regimes.
As for the other considered noise models, the main effect is a reduction  of the first fidelity peak, which nonetheless remains  highly pronounced around $t^{*}=\pi$. This behaviour, consistently observed for all  noise models, highlight an almost universal feature of the routing protocol, effectively promoting $t^{*}\approx\pi$ to a suitable routing time even in the presence of noise.  
In other words, irrespective of the main source of  noise (phases or weight) and of its temporal structure (static or dynamical) the  routing fidelity always reaches its maximum nearly at the same universal time and this is a relevant feature of the scheme.
\section{Discussion}
\label{sec:D}
Chirality constitutes a key resource for the design of efficient quantum routing protocols. 
Particularly, the introduction of chiral phases in the network plays a central role, as it enables interference effects  that cannot be realized  with real Hamiltonians and that are essential for achieving high-fidelity quantum state transfer. Our results  show that the protocol maintains a large degree of    robustness under  different noise conditions suggesting its potential for the development of qubit-based quantum communication schemes. Let us now summarize and discuss results in two specific frameworks.
\subsection{Noisy Qubit Routing}
For routing of qubits, chiral phases are fundamental for achieving perfect quantum routing, as they enforce the directional interference patterns required for deterministic information transfer.  However, when noise perturbs the chiral phases, the routing fidelity decreases with increasing noise strengths with respect to the noiseless case. This degradation is stronger when multiple  output ports $n$ are available. 
This dependence arises because fluctuation in the chiral phases alter the reduced topology of the network (see Fig.\ref{fig:Reduced Lily Graph}) modifying the coefficient $\Gamma(n,\beta,\delta)$ and the connection between the unwanted outputs and the rest of the structure. Phase noise also affects  the coefficient $\Xi(\beta,\gamma)$, which determines the transition probability  from the input to the target output state. 
The second relevant noise  is related is the fluctuation of the weight parameter of the network. Interestingly, the impact of weight fluctuations is independent of the number of outputs. When the chiral phases are correctly set to their optimal values,  the term $\Gamma(n,\beta,\delta=0)=0$ for all the possible value of the parameter $\beta$, ensuring that there is no  leakage of information towards the unwanted outputs. The fluctuation of the $\beta$ parameter, with the ideal values for the phases, only influence the value of the coefficient $\Xi(\beta,\gamma)$ and thus affect solely  the transfer fidelity. 
Our systematic analysis provides a characterization of the routing performance under non-ideal conditions, identifying regimes where high transfer fidelity remains achievable. Remarkably, we have found that, regardless of the noise model or noise strength, the optimal routing time consistently remains $t^{*}\approx\pi$, indicating an almost universal transfer time for noisy quantum routing protocols.
\subsection{Long range noisy qubit routing and qudit routing}
\textcolor{black}{The structure of the Lily network  can be in principle generalized to the routing of qudits, or qubits over arbitrary distant spatial regions, by employing a larger number of nodes. Assuming that  the input and output regions are now composed of $D$ nodes, the router implementation needs a weight distribution as $\tilde H(n,\vec{\beta},\phi_{0},\phi_{1},\phi_{2})$ \cite{cavazzoni2025perfect}. For the correct phase distribution, the routing protocol is effectively reduced to the transport of information from the selected input to the desired output and the noisy time evolution can be compared to the noisy evolution in spin chains \cite{zwick2011robustness,zwick2012spin,stolze2013robustness,nikolopoulos2014quantum}, providing fundamental insights in the resilience of the procedure in engineered couplings. The phase mechanism, corresponding to the chiral and routing layer, is the same as the one presented in Fig.\ref{fig:Lily Graph} promoting the structure presented to the most general mechanism to implement quantum routing in networks. The time needed to perform long range routing procedures turns out to be the same as the standard qubit routing, i.e. $t^{*}=\pi$ \cite{perez2013perfect} ensuring the universality of $t^{*}$ not only in noisy conditions, but also for more complex protocols.}
\section{Conclusions}
\label{sec:C}
We have investigated the performances of quantum routing in a chiral network, focusing on the transfer of arbitrary qubit states  from a given input location to multiple possible outputs. Our analysis have addressed the robustness of the protocol under several noise models, both static and time-dependent thus providing a detailed characterization of quantum routing in noisy networks. 
Our results show that the protocol is robust against moderate noise on the chiral phases or the coupling weight.
For higher noise strengths, the transfer fidelity progressively decreases, indicating the operational limit of the routing scheme. \textcolor{black}{Our results have shown that for networks with a moderate number of outputs ($n \lesssim 30$), noise in the edge weights constitute the main source of degradation, compared to the chiral phase fluctuation.} Moreover, we have identified an almost universal  routing time under any noisy conditions. Indeed, the fidelity consistently reaches its maximum at a time $t^{*} \approx \pi$ in every noisy condition. Overall, our findings offer practical  insight in the design  of realistic  communications architectures and provides guidelines for the realization of next-generation quantum routing networks.


%
%

\ack{This work has been done under the auspices of GNFM-INdAM. The authors thank Luca Casarini for his contribution to the early stage of this project.}

\funding{This work has been partially supported by MUR and EU through the projects PRIN22-PNRR-P202222WBL-QWEST, PRIN22-202224BTFZ-EQWALITY, NQSTI-Spoke1-BaC QBETTER  (contract n. PE00000002-QBETTER), NQSTI-Spoke1-BaC QSynKrono (contract n. PE00000002-QuSynKrono), and NQSTI-Spoke2-BaC QMORE (contract n. PE00000023-QMORE).}

\roles{Conceptualization, S.C., G.R., C.B., P.B. and M.G.A.P; methodology, S.C., C.B., G.R., P.B. and M.G.A.P; software, C.B., G.R. and S.C.; validation, C.B., G.R., S.C., P.B. and M.G.A.P formal analysis, C.B., G.R., S.C., P.B. and M.G.A.P; investigation, G.R. and S.C.; writing---original draft preparation, S.C.; writing---review and editing, C.B., G.R., S.C., P.B. and M.G.A.P; visualization, G.R. and S.C.; supervision, C.B., P.B. and M.G.A.P. All authors have read and agreed to the published version of the manuscript.}

\data{Data can be accessed upon reasonable request.}

\suppdata{No supplementary data were generated for this study.}

\appendix

\section{-- Comparison between Routing and Multiple State Transfer}
\label{app:Comparison}

\begin{figure}[htbp]
  \centering
\begin{subfigure}[b]{0.35\textwidth}
    \raggedleft
\begin{tikzpicture}[scale=1,
  every node/.style={
    rectangle,
    draw,
    thick,
    inner sep=0pt,
    minimum size=8mm
  }
]

\foreach \i in {1,...,6}{
  \node (v\i) at ({60*(\i-1)}:2)
    {\includegraphics[width=5mm]{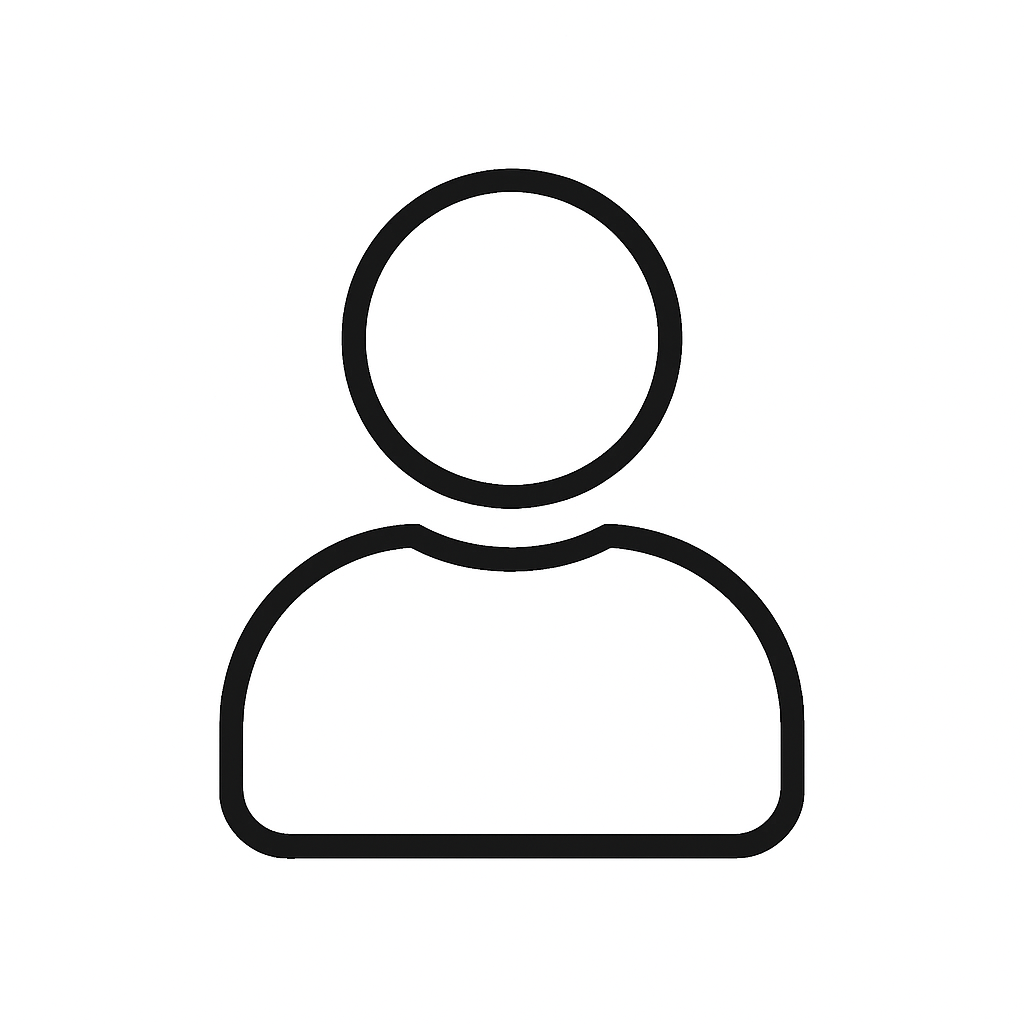}};
}

\foreach \i in {1,...,6}{
  \foreach \j in {1,...,6}{
    \ifnum\i<\j
      \draw[double, dotted] (v\i) -- (v\j);
    \fi
  }
}

\draw[double,thick,red] (v3) -- (v6);

\end{tikzpicture}
\caption{Schematic procedure for multiple state transfer.}
\label{fig:Schematic State Transfer}
\end{subfigure}
  \hfill
\begin{subfigure}[b]{0.35\textwidth}
    \raggedleft
\begin{tikzpicture}[scale=1,
  every node/.style={
    rectangle,
    draw,
    thick,
    inner sep=0pt,
    minimum size=8mm
  }
]
  \node (c) at (0,0) {\includegraphics[width=4mm]{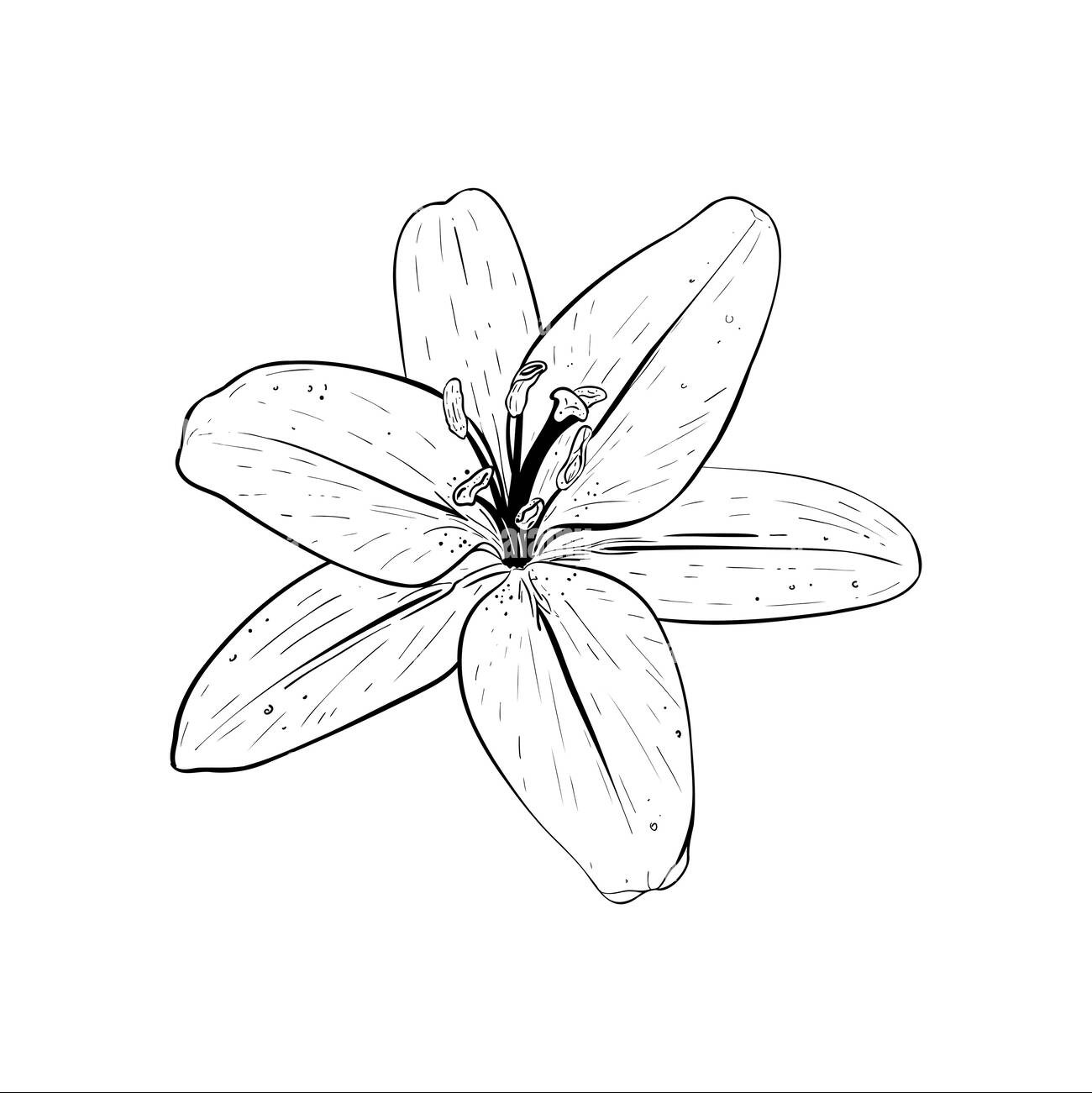}};

  \foreach \i in {1,...,6}{
    \node (v\i) at ({360/6*(\i-1)}:2)
      {\includegraphics[width=5mm]{images/A_simple_black_and_white_vector_illustration_showc.png}};
  }
  
  \foreach \i in {1,...,6} {
    \draw[dotted] (c) -- (v\i);
  }

  \draw[double,thick,red] (v4) -- (c);
  \draw[double,thick,red] (v6) -- (c);

\end{tikzpicture}
\caption{Schematic \textit{Lily Network} topology for quantum routing.}
\label{fig:Schematic Lily Graph}
\end{subfigure}
\caption{(a)-- A perfect state transfer procedure for exchanging information among $n$ senders and receivers requires $n(n+1)/2$ \textit{independent} channels. Each sender activates only one channel (depicted as a continuous double red line) to communicate with the intended receiver, while all other channels (shown as double dotted lines) remain inactive. (b)-- A network connecting among each others $n$ senders and receivers through the \textit{Lily Topology}. The chiral phases select the desired output connecting it to the input (depicted as continuous double red lines) and leaves all the other channels unused.}
\label{fig:State Transfer vs Routing}
\end{figure}

The main distinction between quantum routing and quantum state transfer lies in the different number of  resources required to implement the two schemes.
Consider a system  composed  of $n+1$ senders and receivers that want to exchange quantum information pairwise. Then they need $n(n+1)/2$  {\it independent } channels that support perfect state transfer (see Fig.\ref{fig:Schematic State Transfer}). 
A standard implementation of such channel is a spin chain of length $l_{\rm{QST}}=2D$ \cite{christandl2004perfect}. The total amount of resources $R_{\rm{QST}}$ needed is then 
\begin{equation}
    \label{eq:Resources State Transfer}
    R_{\rm{QST}} = D(n^2+n)
\end{equation}
which scales quadratically with the number of senders and receivers. 

On the other hand, exploiting the $\textit{Lily }$ graph  symmetries (see Fig.\ref{fig:Schematic Lily Graph}) the resources $R_{\rm{QR}}$ required for quantum routing reduce to
\begin{equation}
    \label{eq:Resources Routing}
    R_{\rm{QR}} = D(n+1) + 2,
\end{equation}
since every sender and receiver channel, composed of just $l_{\rm{QR}}=D$ elements, is connected  through the central routing structure (schematically represented as a \textit{Lily} in Fig.\ref{fig:Schematic Lily Graph}). 
Choosing the length $l_{\rm{QR}}=D$ for each routing channel connected to the central  structure, guarantees a fair comparison between the two schemes since the total distance between any sender-receiver pair throught the quantum router matches the  state transfer channel length, i.e. $l_{\rm{QST}}=2D$. The $+2$ contribution in Eq.\eqref{eq:Resources Routing} is due to the central routing structure, which is composed of just two vertices (see Fig.\ref{fig:Lily Graph}) shared by all the senders and receivers.

\section{-- Average routing fidelity  }

\label{app:noise}

In this section we derive in detail expressions for the average routing fidelity under different noise models.  We consider here that the control parameters of the Hamiltonian (namely, the weights and the chiral phases in the pink and blue links of Fig. \ref{fig:Lily Graph}) differ from the optimal ones reported in \cite{cavazzoni2025perfect}, which lead to perfect quantum routing. 
Under the assumptions that the other links (the ones with no weight or phase) are not affected and that the weights $\beta$ fluctuate together, the Hamiltonian can be written in the reduced form of Eq.(\ref{eq:reduced_Hamiltonian_beta}). 
We now analyse two different noise models: static noise, in which the control parameters of the Hamiltonian remain fixed throughout the time evolution, and the dynamic noise, in which they continuously fluctuate in time. 
In both cases we considered fluctuations on either the weight value $\beta$ or on the chiral phases to better distinguish and isolate the effect of the different noises. Regarding the phases, we modelled the noise in such a way that it affects directly the effective phases $\gamma$ and $\delta$ in Eq.(\ref{eq:reduced_Hamiltonian_beta}).

\subsection{Static Noise}

Referring to Fig. \ref{fig:Lily Graph}, we choose as the initial state the most general superposition of the states $\ket 1$ and $\ket 2$, which coincide with the  reduced states $\ket{\tilde 1}$ and $\ket{\tilde 2}$: $\ket{\psi^{\rm{in}}(\alpha,\chi)}=\cos{\left(\frac{\alpha}2\right)} \ket{\tilde 1}+e^{i\chi}\sin{\left(\frac{\alpha}2\right)}\ket{\tilde 2}$. The target state is then weighted with the same coefficients in the chosen output branch: $\ket{\psi^{\rm{out}}(\alpha, \chi)}=\cos{\left(\frac{\alpha}2\right)} \ket{\tilde 5}+e^{i\chi}\sin{\left(\frac{\alpha}2\right)}\ket{\tilde 4}$. In the reduced subspace, the time evolution operator  $\tilde U(n,t, \beta, \gamma, \delta)$ depends on the number of outputs,  time, and the actual configuration of the control parameters (the weight $\beta$ and the effective chiral phases $\{ \gamma, \delta\}$). 
For fixed  Hamiltonian parameters,  the evolution operator is $\tilde U(n,t, \beta, \gamma, \delta)=e^{-i \tilde H(n, \beta, \gamma, \delta)t}$.
Due to the presence of noise, we do not have perfect control over the  parameters $\beta$, $\gamma$ and $\delta$ (denoted from now on by $\vec{\eta}_{H}$), which fluctuate according to a noise model 
described by the set of parameters $\vec{\eta}_n$. The time evolved state is then described by
\begin{equation}
    \label{eq:rho_noisy}
    \rho(\alpha, \chi,n,t, \vec{\eta}_{H}^{\, \rm{opt}}, \vec{\eta}_{n})=
    \int_{\vec{\eta}_H}\cb{\!\!\!d\vec{\eta}_H}\,p(\vec{\eta}_H, \vec{\eta}_{H}^{\, \rm{opt}},\vec{\eta}_{n})\,\tilde U(n,t,\vec{\eta}_{H},\vec{\eta}_n) \ket{\psi^{in}(\alpha, \chi)} \bra{\psi^{in}(\alpha,\chi)} \left({\tilde U}(n,t,\vec{\eta}_{H},\vec{\eta}_n)\right)^{\dagger}.
\end{equation}
The integration is performed over the possible realizations of the Hamiltonian parameters and is controlled by the probability distribution $p(\vec{\eta}_H,\vec{\eta}_{H}^{\, \rm{opt}}, \vec{\eta}_{n})$, which depends on the actual values of the control parameters $\vec{\eta}_H$, on their optimal values $\vec{\eta}_{H}^{\, \rm{opt}}$, and on additional parameters $\vec{\eta}_{n}$ associated to the noise model. 

As we are interested in assessing the routing procedures, the quantity of interest is the fidelity between the evolved state and the target state, which reads 
\begin{equation}
    F(\alpha, \chi,n,t, \vec{\eta}_{H}^{\, \rm{opt}}, \vec{\eta}_{n})=\bra{\psi^{out}(\alpha,\chi)}\rho(\alpha, \chi,n,t, \vec{\eta}_{H}^{\, \rm{opt}}, \vec{\eta}_{n})\ket{\psi^{out}(\alpha,\chi)}.
\end{equation} 

Since there is not a preferred state for routing procedure, to evaluate the overall performances of the structure, we calculate the quantum fidelity averaged over all the possible initial states:
\begin{equation}
    \bar{ F}(n,t, \vec{\eta}_{H}^{\, \rm{opt}}, \vec{\eta}_{n})=\frac1{4\pi}\int_0^{\pi}
    \!\!\!\sin{\alpha}\,d\alpha\int_0^{2\pi}\!\!\!\!\!\!d\chi \,\bra{\psi^{out}(\alpha,\chi)}\rho(\alpha, \chi,n,t, \vec{\eta}_{H}^{\, \rm{opt}}, \vec{\eta}_{n})\ket{\psi^{out}(\alpha,\chi)}.
\end{equation} 
Overall, we have two different integrations: one over the {Hamiltonian parameters} and one over the initial state parameters. By explicitly expressing the evolved state we can rewrite the previous expression:
\begin{align}
    \bar{ F}(n,t, \vec{\eta}_{H}^{\, \rm{opt}}, \vec{\eta}_{n})=\frac1{4\pi}\int_0^{\pi}\!\!\!\!\sin{\alpha}\,d\alpha\int_0^{2\pi}\!\!\!\!\!\!d\chi \, \bra{\psi^{out}(\alpha,\chi)} \bigg[&\int_{\vec{\eta}_H}\cb{\!\!\!d\vec{\eta}_H}\,p(\vec{\eta}_H, \vec{\eta}_{H}^{\, \rm{opt}},\vec{\eta}_{n})\,\tilde U(n,t,\vec{\eta}_{H},\vec{\eta}_n) \ket{\psi^{in}(\alpha, \chi)} \nonumber \times \\ & \bra{\psi^{in}(\alpha,\chi)} \left({\tilde U}(n,t,\vec{\eta}_{H},\vec{\eta}_n)\right)^{\dagger}\bigg]\ket{\psi^{out}(\alpha,\chi)},
\end{align} 
and by changing the order of the integrals

\begin{align}
    \label{eq:mean_fidelity_initial_state_noise}
    \nonumber
    &\bar{ F}(n,t, \vec{\eta}_{H}^{\, \rm{opt}}, \vec{\eta}_{n}) = \nonumber \\
    &=\int_{\vec{\eta}_H}\cb{\!\!\!d\vec{\eta}_H}\,p(\vec{\eta}_H, \vec{\eta}_{H}^{\, \rm{opt}},\vec{\eta}_{n})\bigg[\frac1{4\pi}\int_0^{\pi}\!\!\!\sin{\alpha}\,d\alpha\int_0^{2\pi}\!\!\!d\chi \,\bra{\psi^{out}(\alpha,\chi)}  \,\tilde U(n,t,\vec{\eta}_{H},\vec{\eta}_n) \ket{\psi^{in}(\alpha, \chi)} \times \\ \nonumber &\quad\quad\quad\quad\quad\quad\quad\quad\quad\quad\quad \bra{\psi^{in}(\alpha,\chi)} \left({\tilde U}(n,t,\vec{\eta}_{H},\vec{\eta}_n)\right)^{\dagger}\ket{\psi^{out}(\alpha,\chi)} \bigg]  \\ \nonumber
    &=\int_{\vec{\eta}_H}\cb{\!\!\!d\vec{\eta}_H}\,p(\vec{\eta}_H, \vec{\eta}_{H}^{\, \rm{opt}},\vec{\eta}_{n})\bigg[\frac1{4\pi}\int_0^{\pi}\!\!\!\sin{\alpha}\,d\alpha\int_0^{2\pi}\!\!\!d\chi \, \left|\bra{\psi^{out}(\alpha,\chi)}  \,\tilde U(n,t,\vec{\eta}_{H},\vec{\eta}_n) \ket{\psi^{in}(\alpha, \chi)} \right|^2\bigg]  \\ \nonumber
    &=\int_{\vec{\eta}_H}\cb{\!\!\!d\vec{\eta}_H}\,p(\vec{\eta}_H, \vec{\eta}_{H}^{\, \rm{opt}},\vec{\eta}_{n})\bigg[\frac1{4\pi}\int_0^{\pi}\!\!\!\sin{\alpha}\,d\alpha\int_0^{2\pi}\!\!\!d\chi \, \bigg|  \left(\cos{\left(\frac{\alpha}2\right)} \bra{\tilde 5}+e^{-i\chi}\sin{\left(\frac{\alpha}2\right)}\bra{\tilde 4}\right)  \times \\  &\quad\quad\quad\quad\quad\quad\quad\quad\quad\quad\quad \tilde U(n,t,\vec{\eta}_{H},\vec{\eta}_n) \left(\cos{\left(\frac{\alpha}2\right)} \ket{\tilde 1}+e^{i\chi}\sin{\left(\frac{\alpha}2\right)}\ket{\tilde 2}\right) \bigg|^2  \bigg] .
\end{align}

The integral over all the possible initial states (the quantity in square brackets in Eq. \eqref{eq:mean_fidelity_initial_state_noise}) leads to 
\begin{align}
\mathfrak{F}_{\tilde U} \equiv \left(\left|\tilde U_{51}\right|^2+\left|\tilde U_{42}\right|^2\right) \frac13 + \left( \tilde U_{51}\tilde U_{42}^*+\left|\tilde U_{41}\right|^2+ \left| \tilde U_{52}\right|^2+\tilde U_{42}\tilde U_{51}^*\right) \frac16 ,
\end{align}
which provides a formula for the fidelity averaged over all the possible initial states, given the matrix elements of the evolution operator $\tilde U$. Since we do not have an explicit analytic expression for the evolution operator $\tilde U$, the integral over the Hamiltonian parameters is done numerically:
\begin{align}
\label{eq:average_fidelity}
    \bar{ F}(n,t, \vec{\eta}_{H}^{\, \rm{opt}}, \vec{\eta}_{n})&=\int_{\vec{\eta}_H}\cb{\!\!\!d\vec{\eta}_H}\,p(\vec{\eta}_H, \vec{\eta}_{H}^{\, \rm{opt}},\vec{\eta}_{n})  \, \mathfrak{F}_U(n,t,\vec{\eta}_H,\vec{\eta}_n).
\end{align}
\\
When the noise affects the chiral phases, the integration is over the effective phases $\gamma$ and $\delta$ of Eq.(\ref{eq:reduced_Hamiltonian_beta}), both governed by a von-Mises probability distribution $p(\epsilon)=\frac{e^{k \cos{\epsilon}}}{2 \pi I_0(k)}$ and Eq.\eqref{eq:rho_noisy} becomes:
\begin{align}
     \rho(\alpha, \chi,n,t, \vec{\eta}_{H}^{\, \rm{opt}}, k)=\int_{-\pi}^{\pi}\int_{-\pi}^{\pi}& \frac{e^{k \cos{\epsilon_1}}}{2 \pi I_0(k)} \frac{e^{k \cos{\epsilon_2}}}{2 \pi I_0(k)}
     \tilde U(n,t,\beta =\frac{\sqrt3}2, \gamma=0+\epsilon_1,\delta=\pi+\epsilon_2) \ket{\psi^{in}(\alpha, \chi)} \times\nonumber \\ & \bra{\psi^{in}(\alpha,\chi)} \left(\tilde U(n,t,\beta =\frac{\sqrt3}2, \gamma=0+\epsilon_1,\delta=\pi+\epsilon_2)\right)^{\dagger} d\epsilon_1 d\epsilon_2, 
\end{align}
where $I_0(k)$ is the modified Bessel function of the first kind, and  $\epsilon_1,\epsilon_1$ represent the deviations from the optimal values of $\gamma$ and $\delta$, namely $0$ and $\pi$. The parameter $k$, called concentration parameter, controls the width of the von-Mises probability distribution: the higher it is, the quicker the probability goes to zero when deviating from the mean value. Indeed, for high values of $k$, the von-Mises probability distribution is well approximated by a Gaussian with variance $\sigma^2=\frac1k$. In the end, the averaged fidelity in Eq.(\ref{eq:average_fidelity}) becomes
\begin{align}
\label{eq:average_fidelity_static_phases}
    \bar{ F}(n,t, \vec{\eta}_{H}^{\, \rm{opt}}, k)=\int_{-\pi}^{\pi}\int_{-\pi}^{\pi} \frac{e^{k \cos{\epsilon_1}}}{2 \pi I_0(k)} \frac{e^{k \cos{\epsilon_2}}}{2 \pi I_0(k)} \, \mathfrak{F}_U(n,t,\beta =\frac{\sqrt3}2, \gamma=0+\epsilon_1,\delta=\pi+\epsilon_2)d\epsilon_1 d\epsilon_2.
\end{align}
\newline
If instead the noise affects the value of the weights, the integration is over the values of $\beta$, and is governed by a normal probability distribution with variance $\sigma^2$, Eq.\eqref{eq:rho_noisy} is reduced to:
\begin{align}
     \rho(\alpha, \chi,n,t, \vec{\eta}_{H}^{\, \rm{opt}}, \sigma)=\int_{-\infty}^{\infty}& \frac{e^{\frac{-\zeta^2}{2\sigma^2}}}{\sigma\sqrt{2\pi}}
    \tilde U(n,t, \beta=\frac{\sqrt 3}{2}+\zeta,\gamma=0,\delta=\pi)\ket{\psi^{in}(\alpha, \chi)}  \times \nonumber \\ & \bra{\psi^{in}(\alpha,\chi)} \left(\tilde U(n,t, \beta=\frac{\sqrt 3}{2}+\zeta,\gamma=0,\delta=\pi)\right)^{\dagger} d\zeta, 
\end{align}
where $\zeta$ represents the discrepancy between the optimal value of the weight ($\beta=\frac{\sqrt 3}{2}$) and the actual value. Eq.(\ref{eq:average_fidelity}) becomes then
\begin{align}
\label{eq:average_fidelity_static_weights}
    \bar{ F}(n,t, \vec{\eta}_{H}^{\, \rm{opt}}, \sigma)=\int_{-\infty}^{\infty}& \frac{e^{\frac{-\zeta^2}{2\sigma^2}}}{\sigma\sqrt{2\pi}} \, \mathfrak{F}_U(n,t, \beta=\frac{\sqrt 3}{2}+\zeta,\gamma=0,\delta=\pi)d\zeta.
\end{align}

\subsection{Dynamic Noise}
 Here we assume that the control parameters of the Hamiltonian (either the effective chiral phases $\gamma$ and $\delta$ or the weight $\beta$) continuously fluctuate around their optimal values. We call \textit{realization} $\vec{\eta}_n[\tau]$ the sequence of values that the control parameters assume over time. We numerically generate a realization throughout an Ornstein-Uhlenbeck process centred in the optimal values. 
Once a realization is generated, the evolution operator evolving the initial state at any time $t$ can be obtained in an approximate form  according to the Trotter-Suzuki formula as
\begin{align}
     U(n,t, \vec{\eta}_n[\tau])=e^{-i \tilde H(n,\vec{\eta}_n[t- \Delta t])\, \Delta t }\,e^{-i \tilde H(n,\vec{\eta}_n[t- 2\Delta t])\, \Delta t }...e^{-i \tilde H(n,\vec{\eta}_n[0])\, \Delta t }.
\end{align}
This procedure would give just the evolution operator associated to a single particular realization. The stochastic nature of the OU process requires to mediate the results over the realization of many different possible realizations (enough to arrive at convergence) to obtain reliable physical results. Thus, the evolved state at a certain time $t$ is obtained by averaging the evolved states associated to every realization:
\begin{align}
    \rho(\alpha, \chi, n, t, \vec{\eta}_{H}^{\, \rm{opt}},\theta, \Sigma)=\bigg< U(n,t, \vec{\eta}_n[\tau])\, \rho_{in}(\alpha,\chi) \,\left(U(n,t, \vec{\eta}_n[\tau])\right)^{\dagger}\bigg>_{\vec{\eta}_n},
\end{align}
where $\rho_{in}(\alpha,\chi)=\ket{\psi^{in}(\alpha,\chi)}\bra{\psi^{in}(\alpha,\chi)}$, $\ket{\psi^{in}(\alpha,\chi)}=\cos{\left(\frac{\alpha}2\right)} \ket{\tilde 1}+e^{i\chi}\sin{\left(\frac{\alpha}2\right)}\ket{\tilde 2}$ and we used $\big< \bullet \big>_{\vec{\eta}_n}$ to denote the average over the realization of the process. 
Note that after the averaging procedure, the state depends only on the optimal values of the Hamiltonian parameters ($\vec{\eta}_{H}^{\, \rm{opt}}$: $\beta=\frac{\sqrt{3}}{2}$, $\gamma=0$ and $\delta=\pi$), and on the parameters $\theta, \Sigma$ controlling the stochastic process . As for the static noise case, since we are interested in the routing procedure, the target state $\ket{\psi^{out}(\alpha,\chi)}=\cos{\left(\frac{\alpha}2\right)} \ket{\tilde 5}+e^{i\chi}\sin{\left(\frac{\alpha}2\right)}\ket{\tilde 4}$ is pure, and the quantum fidelity between the time evolved state and the target state reduces to
\begin{equation}
    F(\alpha, \chi,n,t,\vec{\eta}_{H}^{\, \rm{opt}},\theta,\Sigma)=\bra{\psi^{out}(\alpha,\chi)} \,\rho(\alpha, \chi, n, t,\vec{\eta}_{H}^{\, \rm{opt}}, \Sigma) \, \ket{\psi^{out}(\alpha,\chi)}.
\end{equation} 
To asses the mean routing performances and to compare them with Eq.\eqref{eq:average_fidelity_static_phases} and Eq.\eqref{eq:average_fidelity_static_weights} we evaluate the mean value of the fidelity integrating over all the possible initial states:
\begin{align}
    \label{eq:average_fidelity_OU}
    \bar{ F}(n,t,\vec{\eta}_{H}^{\, \rm{opt}}, \theta,\Sigma)&= \frac1{4\pi}\int_0^{\pi}\sin{(\alpha)}d\alpha\int_0^{2\pi}d\chi \,\bra{\psi^{out}(\alpha,\chi)} \,\rho(\alpha, \chi, n, t,\vec{\eta}_{H}^{\, \rm{opt}}, \Sigma) \, \ket{\psi^{out}(\alpha,\chi)} \nonumber \\
    &=\bigg<\frac1{4\pi}\int_0^{\pi}\sin{(\alpha)}d\alpha\int_0^{2\pi}d\chi \, \left| \bra{\psi^{out}(\alpha,\chi)} U(n,t, \vec{\eta}_n[\tau]) \ket{\psi^{in}(\alpha,\chi)} \right|^2 \bigg>_{\vec{\eta}_n} \nonumber \\
    &=\bigg<\mathfrak{F}_U(n,t,\vec{\eta}_n[\tau])\bigg>_{\vec{\eta}_n}.
\end{align}

Practically speaking, since the average over the realizations is just a linear summation, it is possible to exchange it with the integral over the initial states, and the evaluation of Eq.\eqref{eq:average_fidelity_OU} is reduced to the average over the different realizations of Eq.(\ref{eq:average_fidelity}).

\bibliography{bib.bib}

\end{document}